
\magnification=\magstep1
\baselineskip=16pt
\parskip=\smallskipamount
\bigskip

  \def\fr#1/#2{{\textstyle{#1\over#2}}}

{\narrower \narrower

\baselineskip = 13pt

\centerline{{\bf Autobiographical Notes of  a Physicist}}
\bigskip

\centerline{{N. David Mermin}}

\centerline{{Horace White Professor of Physics Emeritus}}

\centerline{{Cornell University}}

\medskip

\bigskip

I describe aspects of my life in physics:  the name I publish under, great physicists I have known, how I got into quantum foundations, what role I've played in it.   My form is autobiographical, but  my personal experience may illustrate what it was like being a physicist over the past 60 years.   I  offer some offbeat ways of thinking about some orthodox physics.

}

\vskip 25pt

My parents named me Nathaniel David Mermin, but they always called me David.   My paternal grandmother, Nechame (Naomi) had died within a year of my birth, and there is a tradition of naming babies after the recently deceased.  Both names start with {\it N\/} and are trisyllabic, with the accent on the second.  Close enough.    I've been told that my parents called me Nathaniel only when visiting my grandfather Shelig (Charles) Mereminskii (Mermin), and that he was aware of their deception.    

I used to call myself David N. Mermin.  Early in my teens I learned that Nathaniel was actually my first name, so I switched to N. David.   After the twin towers came down, airport security started insisting that names on all ID's must agree with names on passports, which must agree with birth certificates.  In the years that followed, the name Nathaniel  drifted onto my driver's license, automobile registration, credit cards, medical appointments, and junk telephone calls.   By 2020 people I didn't know addressed me as Nathaniel or even Nate, until I corrected them.    

My mother Eva was, like most professional women in the 1930s,  a school teacher. She stopped teaching when I was born, and resumed  when my younger brother Joel had left home for college.   My father John was almost 3 years old when his parents were driven by pogroms from Zhitomir, Ukraine, to New Haven, Connecticut.   Charles became a shopkeeper.   

John and his nine younger siblings grew up in New Haven.    He and three of his brothers attended Yale.  His brother Al was chosen to take over the shop.   It being the early years of the 20th century, no higher education was offered to his five sisters, three of whom nevertheless, like Al, built serious careers  for themselves.  John spent the rest of his life in New Haven, as a power engineer at the Southern New England Telephone Company.   

When Charles died in 1946, John, effectively a parent to his youngest siblings,  organized the family circle, which had started to diffuse across the country, into a family Decagon, named for the ten siblings and expressing his love of geometry.   He declared that the Decagon should meet yearly in New Haven, even in the absence of actual parents to draw them back.    He could not have imagined what he was setting in motion.  More than three quarters of a century later the Decagon still meets annually at various conference centers.    There are more than a hundred people.  With the exception of the much younger bride of my father's youngest sibling, its most elderly members are Charles's surviving grandchildren and their spouses.  Born in 1935, I am the oldest living male.  The Decagon is now dominated by Charles's great-grandchildren and great-great grandchildren.

I went to the same high school as my father and his brothers.  Many of my teachers remembered my younger uncles.   I expected to follow them all to Yale.    But at 17 I had grown bored with New Haven.  Harvard was the obvious way out.    

I met the now famous Kenneth G.~Wilson when we were both Harvard freshmen.  He was only 16 and already a wizard at mathematics. I was 17 and just starting to learn calculus, which was not then taught in public high schools.   My roommate, Richard Friedberg, was an even better mathematician than Ken.   I quickly concluded that I could never be a professional mathematician, although I loved the subject.   I had no idea that Dick and Ken were the two most  mathematicially powerful people I would ever meet.  That's what Harvard was like back then.  It  probably still is.

I majored in mathematics in spite of my two formidable friends.   In 1956, to avoid a field in which I could only be mediocre, I applied to graduate school in mathematical statistics.    I was awarded a National Science Foundation predoctoral fellowship to study statistics at Columbia.   But my friend Dorothy Milman had two more years at Radcliffe, and I realized that I didn't  want to leave Cambridge.   Because Harvard was weaker than Columbia in mathematical statistics, I asked the Foundation  if I could change my subject to physics and my university to  Harvard.  ``Sure'', said NSF.   Harvard Physics  required its undergraduate majors to attend graduate school elsewhere, but  since I had been a math major that didn't apply to me.   So I became a physicist and married Dorothy.  
   
As an undergraduate I had taken only one physics course beyond  introductory physics: a course in electromagnetism from the wonderful Edward Purcell.  I began to learn the rest of physics in graduate school.   My first course in quantum mechanics was taught by the great Julian Schwinger.   This saved me from the bizarre introductory course that was still standard in 1956,  beginning with  ``the old quantum theory" and culminating in ``the Bohr-Sommerfeld quantization conditions'', both obsolete subjects that had been replaced by the modern quantum theory 30 years earlier.    

Schwinger's course jumped right into contemporary  quantum theory, but it was his own personal version, based on what he called ``elementary measurement symbols".      Other students viewed this as an off-beat version of Paul Dirac's famous formulation, but I, who had never heard of Dirac, found Schwinger's version mathematically ill-formulated and physically obscure.   I was rescued from pure Schwinger by the best of the few introductory texts, Pauling (Linus) and Wilson (E. Bright, Ken's father), which they had written for their fellow chemists.  It was grounded in simple phenomena, and therefore just right for me.   By figuring out the connection between Schwinger's esoterica and the straightforward clarity of Pauling and Wilson, I acquired an unusual but quite decent grasp of quantum mechanics. I also learned from Schwinger a number of beautiful and powerful mathematical tricks, which served me well during my professional career.  

Schwinger had an unlimited capacity for  supervising Ph.D. theses.   Almost every theory student wanted to work with him.   Fortunately when I was ready to start a thesis, Schwinger's student Paul Martin returned  to Harvard as a faculty member, offering an alternative to spending years waiting in line to see Schwinger outside his office.  This was shortly after Bardeen, Cooper, and Schrieffer had put forth their celebrated  theory of superconductivity, one of the most spectacular achievements of quantum physics.    Paul had the idea that a superficially quite different theoretical model offered an alternative approach to accounting for superconductors.    I soon became convinced that his model did not describe the transition from an ordinary metal to a superconductor.  What it did describe, in a rather crude way, was the condensation of a gas  into a liquid.     My thesis was a record of my efforts to persuade Paul of this.    Many years later at his 60th birthday celebration he declared ``David Mermin's thesis blew up in our faces."  

This thesis that exploded persuaded Paul of his error and surprisingly earned me a PhD.  With that certification I  was awarded an NSF postdoctoral fellowship.  Like many new Harvard theoretical physics PhDs before me, I wrote a letter to Aage Bohr saying that I had a fully paid postdoctoral fellowship and would like to take it at the Institute for Theoretical Physics (later the [Niels] Bohr Institute) in Copenhagen.   Dorothy and I looked forward to two years in a beautiful fairy-tale city.  Unexpectedly Aage Bohr replied apologetically that they were entirely out of office space.   I reported this to Paul. ``Now what?'' I asked.    ``Birmingham!'' said he.  My anticipated fairy-land turned into the dark satanic mills of the English industrial midlands.    Dreams of {\it matjes\/} herring gave way to nightmares of boiled cabbage. 

This  was the most fortunate disappointment  of my life.   All I knew about the University of Birmingham was that Rudolph Peierls was there.  I knew about Peierls because I had stumbled on his book of lectures on solid state physics. It was beautifully written,  much more thoughtful and coherent than any of the few existing texts.   I had no idea that Peierls was an outstanding member of the first generation to apply quantum mechanics to all kinds of physical problems, shortly after Heisenberg and Schr\"odinger had created the general theory in 1925. In reply to my letter of inquiry Peierls said he would be visiting MIT in a few weeks and we should talk then.  He turned out to be as charming as his book.  He said I would be welcome to spend two years at his Department of Mathematical Physics, which, unknown to me,  was one of the world's great centers of theoretical physics, much better fitted to my tastes and interests than Copenhagen.  Birmingham and its environs turned out to be far more interesting to Dorothy and me than Copenhagen would have been.  It is unlikely I would have been offered a faculty position at Cornell two years later, if I had written from Copenhagen.

Mrs. (later Lady) Peierls was was a ferocious Russian, notorious for her determination to reorganize the family lives of all members of the Mathematical Physics Department.  She quickly decided that Dorothy needed no help, and took only a friendly interest in us --- a good thing, since for our first year we lived just around the corner.   Genia told us many years later that Dorothy was the only young wife who had arrived in Birmingham planning to write a doctoral thesis, and actually did.    

While in Birmingham I published a few papers based on my thesis.   Then Peierls dropped by my office to suggest that it was time for me to move on to something else --- an alarming event.    So I wrote a couple of quite different papers with a new Birmingham friend, Eric Canel.    Looking back on everything I published before leaving Birmingham, I see little of much interest.    

In late 1962 I applied for faculty positions.   It was a fine time to be doing this.  Baby-boomers were  reaching college age and universities were expanding to accommodate them.  And the 1957  Soviet launch of Sputnik, the first artificial satellite, persuaded Congress that America had a shortage of scientists.  To my surprise Geoffrey Chester, a young Birmingham faculty member about 5 years older than me, suggested that I write to Cornell, a University a notch or two above the places I was writing.    Cornell had been one of the world's great centers of physics in the early 1950's.   But its two  young luminaries, Richard Feynman and Freeman Dyson, had left Cornell for Caltech and the Institute for Advanced Studies, where they stayed for the rest of their lives, leaving Hans Bethe, in his mid-fifties, as the only world-renowned figure.  Bethe remained at Cornell, doing first-rate physics  for forty more years, right up to his death in Ithaca at the age of 98.  
 
The few publications I extracted from my thesis got me labeled an authority on changes of state.  This made it natural for me to ask for a position in Cornell's Laboratory of Atomic and Solid State Physics (LASSP).    Geoffrey told me that he himself would be starting a tenured faculty position in LASSP.    Furthermore Ken Wilson had himself switched to physics in graduate school and would  be joining the Cornell faculty as a theorist in the Laboratory of Nuclear Studies.    I mailed a letter of inquiry to the Cornell Physics Department, and to my surprise was offered an Assistant Professorship just a few weeks later.   Geoffrey must have sent a strong endorsement, and he must have asked Peierls to write recommending me to Bethe, his old friend and collaborator going back to the early days of quantum physics.  
 
When I told Paul Martin that I had accepted a faculty position at Cornell he asked why I would want to do that, thinking its days of glory long past.   I told him of future glories Geoffrey had promised, but Paul remained unconvinced.   Ten years later, when Cornell had indeed been transformed back into one of the world's great physics departments, Paul was on a visiting committee to evaluate the department.   I ran into him just after he had met with our graduate students.   I asked how it had gone.   ``They think they're in Valhalla," he said.
 
 An important part of this transformation, unknown to any of us in 1962, was a 1965 visit to Ben Widom in Cornell's Chemistry department by a young physicist from England, Michael E. Fisher, a contemporary of Geoffrey.  Michael ended up staying at Cornell for twenty years, as the Horace White Professor of Chemistry, Physics, and Mathematics.    Michael and Ben ran a weekly luncheon, the ``Chemical Physics Seminar'', which played a central role in the development of ``the modern theory of phase transitions" associated with Fisher, Widom, Leo Kadanoff, and Ken Wilson.   The Chemical Physics Seminar, particularly Michael Fisher, was an important part of my intellectual life for two decades.  I had a front-row seat at those historic events, and even played a role in some of them.  
  
Wilson received the 1982 Physics Nobel Prize for this work.  I've always assumed that Alfred Nobel's limit of three people to a prize, combined with the challenge of deciding which of the other three to omit, played a role in  the award going only to Wilson, who had the single most important idea.    Ken and Alison then built a beautiful house in a forest just outside of Ithaca, which Dorothy and I bought from them when they left Cornell in 1988.   We lived there happily for the next 35 years.   But we are leaving Ithaca for Portland Maine in early 2024, to be a short walk away from our son Jonathan.

\bigskip  

As of 2023 my most cited publication by far (almost 10,000 citations --- quite a lot for a physics paper) dates from 1966, and my second most cited (over 3,000 --- still a lot), from 1965.   There are stories behind why both of these are still cited as often as they are.

Shortly after Michael Fisher arrived at Cornell, Herbert Wagner and I tried to explain to him that a then-unpublished argument of Pierre Hohenberg (about superfluids and superconductors)  could easily be adapted to prove that there could be no magnetization in a 2-dimensional version of a theoretical model that Heisenberg had used to describe a ferromagnet.  I hadn't known Michael for very long, and one of the first things I learned from him was that you should think twice before claiming to {\it prove\/} anything to a physicist who took mathematics as seriously as he did.  

He didn't believe a word of it.  It became evident that we were dealing with a man who knew nothing about quantum field theory, didn't care one bit that he didn't, and was convinced that we would be better off ourselves to forget it.  So in response to this astonishing attack we worked backwards, unbundling our result from the conceptual wrappings in which it had been enshrouded by some of the great thinkers of the previous decade.  We peeled off layer after layer, day after day, in the face of Michael's unrelenting skepticism, until finally we had it down to a trivial statement about a finite set of numbers.   

Then  an astonishing change took place. ``Publish!'' he practically shouted, ``It's very important!''     Having learned what it was like to be at the end of a Michael Fisher attack, I learned what it was like to have him on my side.  Freeman Dyson came back for a visit to Ithaca. Michael introduced us. ``Mermin and Wagner have just proved that there is no spontaneous magnetization in the 2-dimensional Heisenberg model,'' Michael proudly informed him, as Herbert and I basked in his admiration. ``Of course there isn't,'' Dyson replied. ``But they have {\it proved\/}  that there isn't,'' Michael insisted. One Dyson eyebrow may have moved up half a millimeter in response.     It would be another 35 years before I managed to impress Freeman Dyson.  No matter.  I had impressed Michael Fisher, and would happily argue with and learn from him for the rest of his life. 

Because there was a current controversy over the Heisenberg model in two dimensions, Wagner and I were able to publish our theorem [1] in the top physics journal dedicated to rapid publication, {\it Physical Review Letters\/}.  Consequently we appeared in print before Hohenberg had published his own theorem that inspired ours.  Although we gave him full credit for the idea, our application of his argument immediately became widely known as the ``Mermin-Wagner Theorem''.   Because the result was called by our names, it was hard to use it without citing our paper.   The argument and its generalizations turned out to have applications in many different areas of physics, whence the enormous number of citations.   All those citations should have been to Hohenberg as well.  

Pierre Hohenberg was a close friend of mine at Harvard in both college and graduate school.   He remained one of my dearest physics friends for the rest of his life.  He had a very distinguished career, and never complained to me that Wagner and I had become famous for an idea that was his.    I have tried to set the record straight.  When the subject comes up in my 1976 book on solid state physics with Neil Ashcroft (about which more below) we say ``the proof is based on an argument of P. C. Hohenberg.''    In the early 21st century the argument is finally starting to be called the ``Hohenberg-Mermin-Wagner" theorem.   It is necessary to keep Mermin and Wagner in the name, so people will know what the longer term refers to.

Pierre also played an important role in my second most cited publication [2], although it is entirely unrelated to my first.     In 1963-1964, between my postdoc at Birmingham and my assistant professorship at Cornell, I spent a year in La Jolla as a postdoc with Walter Kohn.   The Physics Department was just north of San Diego, California, right on a beautiful beach at La Jolla.  I kept a bathing suit and towel in my office.

Shortly after I got to La Jolla,  Walter returned from a sabbatical year in Paris.   When I met him for the first time he told me about a little theorem that he had proved in Paris, working with Pierre.   The proof was one of those clever little three-line arguments that wouldn't have occurred to me if I had thought about it for a hundred years.  It was utterly simple and transparent when Walter laid it out in front of me.   

He asked me to think about how to generalize the theorem from the state of lowest energy (temperature zero) to non-zero temperature.  I went back to my office to think about it and quickly realized that a strange way of thinking about temperature that I had formulated in Birmingham for an utterly unrelated purpose, might be tailor-made for generalizing the Hohenberg-Kohn theorem to non-zero temperature.   It took me less than an hour to check that their proof did indeed work in the same way, even when the temperature was not zero.   
 
So I went back to Walter's office and knocked on the door. ``Here's how you do it,''  I said. He seemed somewhat taken aback, and before I got very far into my explanation, he repeated what it was that he wanted me to work on.   I said yes, that was what I had understood.   My point was that his argument worked just as well when the temperature was not zero, if you thought about temperature in the unfamiliar way I was trying to tell him about. He was deeply skeptical. Slowly it dawned on me that this was the problem he had hired me to spend the year working on.

It took me a day to convince him that I had indeed solved his problem. Then he  was very pleased and I was ecstatic. Throughout childhood I was the last to be picked when baseball teams were being formed. I could never get my bat to make contact with the ball. But one day, by sheer chance, I got the bat in the right place at the right time and the ball went sailing over the heads of the outfielders.  I must have been ten years old. It was a magical moment. Now I was 28, at the beginning of my career in physics, and it had happened again. Never in the sixty years of professional life that followed did I ever have as glorious an experience.

I must note that Walter took a different view of this history. He maintained that it had taken me a day,  not just an hour, to do his year-long postdoctoral project.  But I am sure that doing the job took me an hour.  Another 23 passed before I convinced him that I really had done it.

Since I had already finished the year's work Walter encouraged me to think about whatever I felt like thinking about.   We became good friends. As we said goodbye at the end of my stay he said, ``By the way, when you get to Cornell, write up that theorem of yours.   Some day it may be important.''  I wasn't so sure, but I did think that he ought to get something out of having  maintained me in that semi-tropical paradise for almost a year.  So I dutifully wrote a short paper in Ithaca.

I sent a draft to Pierre. He phoned. Neither of us thought this was important. ``If Walter says so then you should publish, but keep it very short.''   I did.    For many years my paper was cited only by Walter.  After half a dozen years, other citations started to creep in.   In the eighties and early nineties they averaged about 20 a year.   Chemists were finding it interesting. Eventually Walter won the 1998 Nobel Prize {\it in chemistry\/} for his paper with Pierre and his other papers that grew out of it.     At that point my own citations started to skyrocket.   Chemists tend to cite more papers than physicists do.  With 3,000 citations, mine may well be one of the most cited two-page papers in the history of physics.  Not bad for an hour's work.  

My third most cited paper [3] is more recent (1992), and is threatening to overtake \#2.  It was written with Charles Bennett and Gilles Brassard.  A paper by  Artur Ekert argued that a famous theorem of John S. Bell provided a basis for an unbreakable cryptography that exploited quantum physics.  I read his paper and realized that Bell's Theorem had nothing to do with it.   But the cryptographic aspect remained valid and was interesting.   I wrote a short paper straightening out Ekert's argument.  I knew that Bennett and Brassard had also written about using quantum mechanics to make a cryptographic device, so I sent my manuscript  to Bennett, whom I had known for years, asking whether he could take a quick look and tell me whether I'd said anything foolish. 

The phone rang early one morning, waking me up.  It was Charlie Bennett, excited by what he viewed as my extension of his work with Brassard.   He was calling to say that we three should write a joint paper.  So he and Brassard wrote a rather long paper.   I cut it down to a rather short paper.  They expanded it back to a less long paper, which I cut down to a less short paper.    After a while I said we'd never agree on what was and was not important to say, so we should each write our own papers.   But Charlie insisted that the three of us should write a joint paper.  What emerged was an uneasy compromise.    I never understood why it attracted so much attention.    For me the most important thing about our paper is that it introduced me to Gilles Brassard. We became friends, visiting each other several times in both Montreal and Ithaca.

The fourth paper (2,300) on my list of most cited is the first that reports the work of more than a day or two.  It's a  56-page pedagogical treatise [4].  Writing it filled an entire sabbatical year (1978-9).  It explains how a branch of mathematics called ``homotopy theory'' can be used to classify imperfections in otherwise ideal  materials.   For many years my review was the only exposition intelligible to physicists, because I knew only a little about homotopy theory and assumed  that my reader knew nothing whatever.   Some physicists wrongly credit me with having developed the whole subject.  

\bigskip 
 
 The year Geoffrey and I arrived at Cornell so did John Wilkins, as an Assistant Professor, like me.  We shared an office and learned a lot from each other.    Geoffrey and John had earlier met at Cambridge, while examining a graduate student named Neil Ashcroft.   At Cornell they enthusiastically recommended Neil for a postdoctoral position, where he arrived the year after Geoffrey, John and I did. After a single postdoctoral year he was himself appointed an Assistant Professor and, like Geoffrey and me, he remained a member of the Cornell Physics Department for the rest of his life.  To my great regret Wilkins left us for Ohio State in 1986.  He lured Ken Wilson there two years later, thereby playing a crucial role in how Dorothy and I acquired our beautiful house in the forest.

Neil and I became close friends soon after his arrival. In 1966 Ashcroft, Wilkins, and I decided to write a textbook on solid state physics.   We called the book Wilkins, Ashcroft, and Mermin: WAM (pronounced ``Wham!''). We started to produce first drafts of chapters in 1967. It soon became clear that John,  the most prolific educator of PhDs in the department, was far too strongly engaged in his research to devote a serious effort to writing a textbook.   Neil and I suggested he withdraw from the collaboration.  I believe John was as relieved as we were.  WAM shrank to AM.  

The happiest years of my professional life were 1968-76 when Neil and I wrote our book and saw it into print. We named it {\it Solid State Physics\/} [5], but today it is known throughout the scientific world as {\it Ashcroft \& Mermin.\/}  Neil was fascinated with materials. Each was like a personal friend. I had little interest in or knowledge of particular materials, but I was fascinated by the conceptual structure that encompassed them all.  I regard our book as an 800 page record of Neil's efforts to teach me solid state physics.  There is a hint of this in the last paragraph of our Preface. I thank Rudolf Peierls for having ``converted [me] to the view that solid state physics is a discipline of beauty, clarity, and coherence''. Then Neil adds that ``having learnt the subject from John Ziman and Brian Pippard, [he] has never been in need of conversion.'' 

At Harvard, solid state physics was not even taught in the Physics Department; it was relegated to the College of Engineering. So I learned nothing about the physics of solids until I was a postdoc at Birmingham and attended Peierls' beautiful lectures in 1961-63.  My meager knowledge of the field made me a bit nervous in 1964, when I started my faculty job at Cornell's Laboratory of Atomic and Solid State Physics.   I taught solid-state physics the following year, as an exercise in self-education.   Fortunately Neil arrived and I immediately began learning from him.  

The book occupied about half our time for eight years. Neil wrote most of the first drafts. I would rarely understand what general issue he was addressing and would revise his essay into something broader that made better sense to me. Neil would then correct whatever misconceptions I had introduced. Back and forth we went, slowly converging on something that looked good to us both. That was before the age of personal computers and text editors. I typed every page on a state-of-the-art IBM ``bouncing-ball'' typewriter, making revisions with a white ``erasing ribbon'' and redoing entire pages when changes were more than a few words. It was a slow process. With today's writing technology it would have taken us less than half the time.   

Neil had a fine sense of humor. He was an excellent mimic and did a superb Hans Bethe. We had  fun dealing with each other's idiosyncrasies throughout the process, and our fun permeates the book.  In 1990 I remarked to Hans that {\it Ashcroft \& Mermin\/} was 14 years old but still in its first edition. He said this showed ``the stability of the subject.''   True enough.   But that's a necessary, not a sufficient condition. Today{\it A\&M\/} is still being used in its first edition, nearly half a century after it first appeared.  Another reason for this longevity is that, unlike most  technical books, ours entertains the reader, just as Neil and I entertained each other during our six years of writing.  

We even had fun reading page proofs and putting together our enormous index, not an easy job in the 1970s. Every entry was written by hand on a ``3-by-5 [inch] card''. If we stacked them, the pile would have been a couple of meters high. Neil's favorite index entry is ``Cart, before horse, 92,'' followed nine pages later by ``Horse, after cart, 92.'' My own favorite (on page 808) is ``Exclamation marks, 61, 185, 219, 224 (twice!), 291, 305, 403, 808.''

The original publisher of our book, long ago swallowed up by bigger and bigger conglomerates, was so eager for us to sign their contract that they offered us terms unusually lucrative half a century ago, and entirely unimaginable today.   This gave us a financial disincentive for producing a new edition with a new contract, as long as people were still buying our first edition.  Neither of us could have imagined fifty years ago that our first edition would outlive Neil, into his eighties, still selling over a thousand copies a year.  Our 1976 first edition has now been translated into Russian (1979), Japanese (1981-2), Polish (1986), German (2001), French (2002), and Portuguese (2011).  

Neil and I had only two other collaborations. One was  a memorial article, ``Hans Bethe's Contributions to Solid-State Physics'' in 2006, the centenary of Bethe's birth; the other (with Malvin Kalos) was an obituary of Geoffrey Chester in 2014. Our revisions and re-revisions were unbelievably easier in the modern era, but were just as extensive. It was clear that in spite of the advances in literary technology, we no longer had the energy to produce a new edition of our book, even had we thought one was needed.

 \bigskip

Looking back on my more than sixty years in physics since my PhD, I realize that my career has been as much literary as scientific.   In the mid 1980s, Gloria Lubkin, editor of {\it Physics Today\/}, the monthly magazine sent to all members of the American Physical Society,  invited me to contribute to a new column of opinion called  ``Reference Frame''.  Earlier that decade I had published two articles in {\it Physics Today\/}.  The first described my successful effort to make the ridiculous word ``boojum'' an internationally accepted scientific term. The second gave a very elementary way to think about Bell's Theorem and its implications for understanding quantum physics.   One was pure entertainment.   The other was a major way to make accessible to nonscientists a centrally important piece of quantum physics.   These  essays persuaded Gloria that I'd make a good columnist.  I wasn't sure. Having to produce something clever and entertaining at regular intervals was not my style. On the occasions when I'd managed to do it, it seemed like a small miracle, unlikely to happen again. So while I didn't say no, I kept stalling. A couple of years went by.   
 
 Then one day I discovered that {\it Physical Review Letters\/}, the world's most widely read
physics journal, was doing something utterly ridiculous that seemed to have escaped the
attention of every physicist I mentioned it to. The absurd policy, and my theory of why
nobody had noticed it, made a good story [6].    Another miracle.   I sent the story to Gloria and became a columnist in 1988, joining a group of Reference Frame writers that included such eminent physicists as Phil Anderson, David Gross, Leo Kadanoff, Dan Kleppner, Jim Langer, and Frank Wilczek.  For a quarter of a century the Reference Frame columns were the most interesting and entertaining part of {\it Physics Today\/}.

After my debut Gloria would phone every few months requesting more miracles. Somehow
she managed to induce them. I came to regard her as my Muse. For over two decades  she extracted essays I didn't know were in me.   She criticized first drafts and negotiated final versions.  My relations with editors have often been tense, but working with Gloria was always a pleasure. 

My most important contribution to science may well have been my May 1991 column, ``Publishing in Computopia'' [7].  I criticized the prevailing culture of paper ``preprints'' and advocated a publicly available  ``electronic bulletin board'' on which anyone could post or read a paper.   Paul Ginsparg has told me that this inspired him to establish three months later at Los Alamos what evolved over the next decade into arXiv.org at Cornell, today one of the great avenues of international scientific communication.   Paul had been thinking about such an undertaking before my column appeared, but my plea (``Why do we live this way?  What are we waiting for?'') stirred him into action.  

In 2009 Gloria Lubkin retired and the Reference Frame columns sadly came to an end.   I found to my surprise that I had produced thirty of them.   Not all were miracles, but several were.  As I traveled around the world of physics after 1988, giving technical talks at universities and conferences, I discovered that I was better known for these essays in {\it Physics Today\/} than for my scientific papers or books.

I had invented an alter ego,  Professor Mozart (``W.~A.'').   I would discuss with him contentious issues, often assigning some of my more outrageous views to W.~A., trying unsuccessfully to talk him out of them.    Many physicists thought Professor Mozart was Neil Ashcroft (whose initials, now that I think of it, were N.~W.~A.).  I refused to say who Mozart actually was to those who mistook my literary flourish for a real person.  Neil told me that as a speaker he was once introduced as ``Neil Ashcroft, better known as Professor Mozart.''  He said nothing to challenge that error.

In 2016, ten years into retirement, I published all thirty of my Reference Frame columns and several related essays in {\it Why Quark Rhymes with Pork and Other Scientific Diversions\/} [8].  This title was a disaster.   It led bookstores to shelve the book not with popularizations of science, as they did with my 1990 collection {\it Boojums All the Way Through\/}, but with technical treatises on high energy physics.   Cambridge University Press should have warned me not to put ``Quark'' in the title.    It did save the book from being displayed with cookbooks (``Pork'').  But they should have made me name it something like {\it Entertainments in  Physics\/}.

I have published about 140 technical physics papers over a 63 year span from 1959 to 2022.  A little over two a year --- not many for a contemporary theorist.  I would characterize most of them as clarifications, refinements, or illustrations of existing theories.   About half of them were published jointly with my graduate students.   Typically I would include a student as a coauthor of the paper that I expected to evolve into their thesis.  I put my name on one or two additional papers only if I had made major contributions.    I asked my students to write  all subsequent papers as sole authors.   This enabled them to learn how to write technical papers, and it helped them find jobs when they finished their PhD. 

My pedagogical papers, some addressed to the general reader, some to physicists, and some  to both groups, have given me the greatest pleasure and are most widely known in the scientific community.   The two awards of which I am most proud are both personal  letters about pedagogical articles, received unexpectedly in the (paper) mail, from the two physicists I admired most in the world.     The first (dated my birthday!) was from Richard Feynman:
\vskip 6pt
 
{

\narrower\narrower

\parindent 0pt
\parskip 8pt
\baselineskip 13pt

California Institute of Technology
\vskip -8pt
Charles C. Lauritsen Laboratory of High Energy Physics
\vskip -8pt

March 30, 1984

Dear Dr. Mermin,

One of the most beautiful papers in physics that I know of is yours in the American Journal of Physics {\bf 49} (1981) 10.

All of my mature life I have been trying to distill the strangeness of quantum mechanics into simpler and simpler circumstances.  I have given many lectures of increasing simplicity and purity.  I was recently very close to your description (down to six states, instead of three, etc.) when your ideally pristine presentation appeared.

I have since copied it almost exactly (with attribution of course) in several recent lectures on the subject.  Thank you.

I have been making a similar series of attempts to explain the relation between spin and statistics.  Can you do as well there?  Perhaps if we meet some day we can discuss it together and create a clear explanation of why exchanging two particles implies a tacit rotation of the axes of one by $360^0$ relative to the other.

                                                Sincerely,

                                                  Richard P. Feynman
                                                 
  }                                                                                                  
 
 \vskip 5pt                                                                                                  
 \noindent The second letter came nearly 7 years later, from Freeman Dyson: \vskip 3pt

{

\narrower\narrower

\parindent 0pt
\parskip 8pt
\baselineskip 13pt

The Institute for Advanced Study

\vskip -8pt
School of Natural Sciences

\vskip -8pt
December 31, 1990

Dear David Mermin,

Thank you for your beautiful paper on the hidden variable theorems which arrived as a New Year gift in today's PRL [{\it Physical Review Letters\/}].  This finally heals the rift between the fans of Kochen-Specker (to which I belong) and the fans of Bell.  This has been for 23 years a dialog of the deaf.  Now, thanks to you, we can speak the same language and understand one another.

Happy New Year!

                                                    Yours,

                                                     Freeman Dyson
                                                     
                                                     }
                                                     
 \vskip 10pt    
 
 \noindent Both letters are about articles on the meaning of quantum mechanics, an interest I acquired in 1981 while teaching an introductory course.    Feynman wrote about an essay  [9] addressed to the general reader, while Dyson wrote about a slightly more technical essay [10] addressed to physicists. 
 
Feynman's 1984 letter provides evidence, two decades after John Bell's famous 1964 paper, that Feynman himself knew nothing about Bell's Theorem, and had figured out the argument entirely on his own.    Indeed interest in quantum foundations was quite low through the early 1980s.   I only learned of Bell's work shortly  before I wrote the 1981 paper that pleased Feynman.  I had learned abut Bell from Max Jammer's book on the philosophy of quantum mechanics.  According to Jammer, that version of Bell's Theorem was discovered by Richard Friedberg, who only learned of Bell's prior work when he showed Jammer his own argument.   Feynman was a master at simplifying physical arguments, so I was glad that he thought I had published an argument even simpler than his own.  But credit for this very simple version of Bell's Theorem belongs (via Jammer's book) to Dick Friedberg, who never published his own argument after he learned that Bell had already done it.

In his final paragraph Feynman offers a characteristically offbeat way of thinking about a fundamental phenomenon.   Although we had met before (as he does not remember) I never saw him again.

To understand Dyson's letter, note that quantum mechanics deals with quantitative properties of a physical system.   It treats the value of such a property as a number displayed by a second physical system, called a measurement apparatus, that has an appropriately designed interaction with the first physical system.    Prior to quantum mechanics it was taken for granted that the numerical values of its properties were inherent in the first physical system, and were merely {\it revealed\/} by an appropriately designed measurement apparatus.  But quantum mechanics maintains that such properties cannot be assigned to the system alone, but can only be associated with the system {\it together with\/}  the particular apparatus used to  display them.   

A natural question arises:  Is this inseparability of a property from the apparatus that displays its value merely a philosophical matter?   Or can it be deduced directly from the specific values that the quantum theory assigns to appropriately chosen sets of properties and measurements?  Strangely, quantum physicists did not ask this basic question until the mid 1960s, four decades after the formulation of their subject.    Two apparently quite different ways were then discovered to deduce the inseparability from specific numerical data.   My 1990 paper pleased Dyson by giving a simple connection between the two ways, thereby healing the quarter-century rift between those who preferred one version to the other.

These letters from Feynman and Dyson have hung framed in my Cornell Physics Department office for over thirty years.   I have received no higher honors.  

\vskip 10pt
\centerline{ --------------------- }
\vskip 20pt

[1] N.~D.~Mermin and H.~Wagner  ``Absence of Ferromagnetism or Antiferromagnetism in One- or Two-Dimensional Isotropic Heisenberg Models", Phys.~Rev.~Lett.~{\bf 17},  1133 (1966); 
inspired by P.~C.~Hohenberg, ``Existence of Long-Range Order in One and Two Dimensions", Phys.~Rev.~{\bf 158}, 383 (1967).

[2] N.~D.~Mermin, ``Thermal Properties of the Inhomogeneous Electron Gas", Phys.\ Rev.\ A {\bf 137}, 1441 (1965).

[3] C.~H.~Bennett, G.~Brassard, N.~D.~Mermin, ``Quantum Cryptography Without Bell's Theorem", Phys.~Rev.~Lett.~ {\bf 68}, 557 (1992).

[4] N.~D.~Mermin, ``The Topological Theory of Defects in Ordered Media", Revs.\ Mod.\ Phys.\ {\bf 51}, 591 (1979).

[5] N.~W.~Ashcroft and N.~D.~Mermin, ``Solid State Physics'', 1st Edition, Saunders College Publishing (1976).

[6] N.~D.~Mermin, ``What's Wrong With This Lagrangean?", Physics Today {\bf 41} (4), 9-11 (1988).

[7] N.~D.~Mermin, ``Publishing in Computopia", Physics Today {\bf 44} (5), 9-10 (1991).

[8] N.~D.~Mermin, ``Why Quark Rhymes with Pork: And Other Scientific Diversions'', Cambridge University Press (2016).

[9] N.~D.~Mermin, ``Bringing Home The Atomic World: Quantum Mysteries for Anybody", American Journal of Physics {\bf  49}, 940 (1981).

[10] N.~D.~Mermin, ``Simple Unified Form for The Major No-Hidden-Variables Theorems", Phys.~Rev.~Lett.~{\bf 65}, 3373  (1990).

 \bye